# Single Experimental Setup for High Sensitive Absorption Coefficient and Optical Nonlinearities Measurements

A. Sharan, S. Abbas Hosseini, D. Goswami*

Tata Institute of Fundamental Research, Homi Bhabha Road, Mumbai 400 005, India.

debu@tifr.res.in

Accurate knowledge of absorption coefficient of a sample is a prerequisite for measuring the third order optical nonlinearity of materials, which can be a serious limitation for unknown samples. We introduce a method, which measures both the absorption coefficient and the third order optical nonlinearity of materials with high sensitivity in a single experimental arrangement. We use a dual-beam pump-probe experiment and conventional single-beam z-scan under different conditions to achieve this goal. We also demonstrate a counterintuitive coupling of the non-interacting probe-beam with the pump-beam in pump-probe z-scan experiment.

* corresponding author

## I. Introduction

Development of high power laser sources has motivated an extensive research in the study of nonlinear optical properties and optical limiting behavior of materials. There exits a continued effort in making sensitive measurements on absorption coefficient and nonlinear coefficients, however, most of the experimental techniques are focused on measuring one or the other of these two important parameters. A variety of interferometric methods[1,2], degenerate four wave mixing[3] , nearly degenerate three wave mixing[4] and beam distortion measurement[5], have been used for measuring the nonlinear refractive index. One of the most important techniques to measure nonlinear refractive index was shown by Sheik Bahaei et al.[6] This technique is simple and versatile yet is highly sensitive. However, an accurate knowledge of absorption coefficient ($\alpha_0$) is necessary for the use of this technique, which is a serious limitation for unknown samples. An easy way to measure $\alpha_0$ is to use the Beer's law[7-8], which operates in the linear absorption regime and has limited sensitivity. More sensitive methods have been developed, of which the technique[9-15] using thermal lens (TL) effect is perhaps the simplest and the most effective. In this method, a lens focuses the laser beam into the sample resulting in a temperature gradient, which in turn produces a spatial gradient in refractive index. The relative change in transmittance of the laser beam can then be measured, after passing through an aperture, with the help of a detector[9-11]. Shen et al.[12-14] introduced a pump probe laser scheme under mode-mismatched and mode-matched conditions to improve sensitivity of the TL method. More recently, Marcano et. al.[15] have used this method to measure the absorption coefficient of water with high accuracy.

A single experimental technique to measure both the parameters, however, is yet to emerge, which could be of significance in the study of new materials. In this paper we introduce a single experimental technique to measure both the parameters. Our aim has been to measure the absorption coefficient ($\alpha_0$) as well as the real and imaginary parts of third-order optical nonlinearity ( $\chi^{(3)}$ ) with high sensitivity in a transparent sample using a single experimental setup. We also show how mid-IR absorption in water (at 1560nm) manifests itself as a minute nonlinear absorption coefficient in the near-IR transmission window (at 780nm). Such high sensitive experiments have become possible due to the ultrahigh sensitivity provided by a stable ultrafast laser operating at both the above wavelengths.

Our technique is a modification of the well-known z-scan technique introduced by Shiek Bahaei et. al[6] in 1990, where one measures the change in transmittance of a focused laser beam through sample that is being moved through the focal point of the lens. Since we are dealing with Gaussian optics when the beam is passing through lens and sample, we will use the Gaussian optics formalism. Subsequently, we will discuss the z-scan theory and finally we will discuss the modifications that we have introduced in addition to the new experimental results and discussions.

## II. Background

### a. The Gaussian beam in a homogeneous medium

In most laser application it is necessary to focus, modify or shape the laser beam by using lenses or other optical components. In general, laser beam propagation can be approximated by assuming that the laser beam has an ideal Gaussian intensity profile corresponding to $TEM_{00}$ mode. Using Maxwell equation in an isotropic, charge free medium one can derive the wave equation[16,17]:

$$\nabla^2 E + k^2 E = 0 \tag{1}$$

where $k^2 = \mu_0 \varepsilon_0 \omega^2 = 2\pi / \lambda$. Let us assume a solution whose transverse dependence is only on $r = \sqrt{x^2 + y^2}$, which will enable us to replace $\nabla^2$ by $\nabla^2{}_t + \frac{\partial^2}{\partial z^2}$ in Eq.(1). We consider nearly plane wave situation where the flow of energy is along a single direction (e.g. z), and therefore the electric field, *E*, is:

$$E = \Psi(x, y, z) e^{-ikz} \tag{2}$$

Substituting these in Eq.(1) we derive

$$\nabla^2{}_t \Psi - 2ik \frac{\partial \Psi}{\partial z} = 0 \tag{3}$$

where we have assumed that longitudinal variation is slow, such that $k \frac{\partial \Psi}{\partial z} \geq \frac{\partial^2 \Psi}{\partial z^2} \leq k^2 \Psi$ is valid. In the next step, we take $\Psi$ of the form

$$\Psi = \exp\{-i[P(z) + \frac{1}{2} Q(z) r^2]\} \tag{4}$$

By substituting Eq.(4) in Eq.(3) we derive

$$-Q^2 r^2 - 2iQ - kr^2 Q' - 2kP' = 0 \tag{5}$$

If this equation is to hold true for all *r*, then the coefficients of different powers of *r* must be equal to zero, which leads to:

$$\begin{aligned} &Q^2 + kQ' = 0 \\ &P' = -i \frac{Q}{k} \end{aligned} \tag{6}$$

For solving this differential equation we introduce a function *S(z)*, such that

$$Q = k \frac{S'}{S} \tag{7}$$

Replacing the value of *Q* in Eq.(6) with the relation from Eq.(7), we get

$$\left(\frac{kS'}{S}\right)^2 + k\left[\frac{kS''S - S'^2}{S^2}\right] = 0 \tag{8}$$

which implies $S'' = 0$, and consequently,

$$S' = a \text{ and } S = az + b \tag{9}$$

where $a$ and $b$ are arbitrary constant. Replacing the values from Eq.(9) in Eq.(7), we get

$$Q = k\frac{a}{az+b} \tag{10}$$

It is more convenient to deal with a parameter $q$, where $q(z) = \frac{k}{Q(z)}$. So that we can rewrite Eq.(9) as:

$$q = z + q_0 \tag{11}$$

where $q_0$ is a constant ($q_0 = b/a$). From Eq.(6) and (11) we have

$$P' = \frac{-i}{q} = \frac{-i}{z+q_0} \quad \Rightarrow \quad P(z) = -i\ln(1+\frac{z}{q_0}) \tag{12}$$

where the arbitrary constant of integration is chosen as zero. The constant of integration will modify the phase of the field solution, Eq.(2). Since the time origin is arbitrary, the phase can be taken as zero. Combining Eqs.(11) and (12) in Eq.(4), we obtain

$$\Psi = \exp\{-i[-i\ln(1+\frac{z}{q_0} + \frac{k}{2(q_0+z)}r^2]\}. \tag{13}$$

We take the $q_0$ to be purely imaginary and express in terms of new constant $w_0$ as $q_0 = i\frac{\pi w_0{}^2}{\lambda}$. By substituting $q_0$ in Eq.(13) and defining following parameters:

$$w^2(z) = w_0{}^2[1+(\frac{\lambda z}{\pi w_0{}^2})^2] = w_0{}^2(1+\frac{z^2}{z_0{}^2}) \tag{14}$$

$$R(z) = z[1+(\frac{\pi w_0{}^2}{\lambda z})^2] = z(1+\frac{z_0{}^2}{z^2}) \tag{15}$$

$$\eta(z) = \tan^{-1}(\frac{\lambda z}{\pi w_0{}^2}) = \tan^{-1}(\frac{z}{z_0}) \tag{16}$$

where $z_0 = \frac{\pi w_0{}^2}{\lambda}$. We can write Eq.(2) as:

$$E = E_0\frac{w_0}{w(z)}\exp\{-i[kz-\eta(z)] - r^2[\frac{1}{w^2(z)} + \frac{ik}{2R(z)}]\} \tag{17}$$

and we can also write

$$\frac{1}{q(z)} = \frac{1}{R(z)} - i\frac{\lambda}{n\pi w^2(z)} \tag{18}$$

which is the fundamental Gaussian beam solution. The parameters $w(z), w_0$ are beam spot size and minimum spot size at *z=0* and the parameter *R(z)* is the radius of curvature of the spherical wavefronts at $z$. Our aim is to calculate spot size of the beam when it passes through a thin lens of focal length *f* as shown in the Fig.1. Since, at the input plane (1) of Fig. 1, $w = w_{01}$ and $R_1 = \infty$, we can write using Eq.(18) the following relation:

$$\frac{1}{q_1} = \frac{1}{R_1} - i\frac{\lambda}{\pi w_{01}{}^2} = -i\frac{\lambda}{\pi w_{01}{}^2} \quad (19)$$

Similarly, at the output plane (2) of Fig.1, we get:

$$\frac{1}{q_2} = \frac{1}{q_1} - \frac{1}{f} = -i\frac{\lambda}{\pi w_{01}{}^2} - \frac{1}{f} \quad (20)$$

or

$$q_2 = \frac{1}{-\frac{1}{f} - i\frac{\lambda}{\pi w_{01}^2}} \quad (21)$$

Finally, at plane (3), $q_3 = q_2 + l$, and the output beam waist, $R_3 = \infty$. Thus, the location of new waist is

$$l = \frac{f}{1 + \left(\frac{f}{z_0}\right)^2} \quad (22)$$

and the minimum spot size in focal point is equal to:

$$w_{03} = w_{01}\frac{\frac{f}{z_0}}{\sqrt{1 + \left(\frac{f}{z_0}\right)^2}} \quad (23)$$

where $z_0 = \frac{\pi w_{01}{}^2}{\lambda}$. The other parameter of interest is the Rayleigh range (*RR*), which is the axial distance from the point of minimum beam waist ($w_0$) to the point where the beam diameter has increased to $\sqrt{2}\,w_0$ in the region of a Gaussian beam focus by a diffraction-limited lens. This is given by the expression:

$$RR = \frac{\pi w_0^2}{\lambda} \quad (24)$$

We are using a lens with focal length *f*=75cm and $w_{01}$=2.77 mm (which is measured by integrating the residual intensity that is measured by translating a knife edge across the beam (Fig. 2). With this background on Gaussian optics, we now discuss the technique of Sheik Bahei et. al.[6] in the following section.

### b. The Z-scan technique

The technique introduced by Bahaei et. al[6] is now popularly known as the z-scan technique[18] as it involves the motion of the sample in the sample across the focal point of laser beam along the direction of propagation of the laser beam (Fig. 3). Assuming

Gaussian beam optics as discussed in the previous section, this experiment allows an intensity scan of the irradiated sample, and provides information about the nonlinearity in the sample. The typical z-scan is performed by translating the sample along the z axis from one side of the focus to the other (fig.3). This results in changing the spot size of the incident beam on the sample to a minimum at the focus and then increasing again on crossing the focus. Correspondingly, the intensity of incident light increases on approaching the focus till a maximum at the focus is reached and then reduces on moving away from the focus. Thus, the overall purpose of the experiment is to determine the variation in transmission as the incident intensity changes by translation along the z-axis. The change in the transmittance of the focusing Gaussian beam in a medium is recorded as a function of position of medium. The transmitted beam is collected either completely (which is called the open aperture case) or through a finite aperture (A) as shown in Fig.3.

Let us first discuss schematically, a simple case of a thin sample with negative nonlinear refractive index when the aperture is closed (A=0.5, which means just 50% of the beam passes through the aperture). When it moves in the z direction it can act as a thin lens with variable focal length. If we start the scan from –z (far from focal length), where the nonlinear refraction is negligible, the transmittance remains relatively constant. As the sample moves closer to the focus, the beam irradiance increases because of self-focusing of the beam will tend to collimate the beam and cause a beam narrowing at the aperture which results in an increase in the measured transmittance (fig. 4a). As the scan continues and sample passes the focal plane, the self-defocusing phenomena will occur. This will broaden the beam at the aperture and a corresponding decrease in transmittance will continue until the sample reaches +z (that is sufficiently far from focus) such that the transmittance becomes linear.

If we open the aperture (A=1) and do the same scan again from –z direction, the transmittance will increase till focal point and as discussed above, it will decrease to the linear case when the sample moves away from focal point to the +z direction (Fig.4c).

Thus, the open aperture case scan gives information on purely absorption nonlinearity while a close aperture case scan contains information about the absorption and dispersion nonlinearity. In case of materials with positive refractive index the story is the reverse of the above cases (Fig.4b and Fig.4d). Induced beam focusing and defocusing of this type have been observed during nonlinear refractive measurement of some semiconductors[10,11].

Let us now consider the above qualitative discussion mathematically. We consider a sample with third order nonlinearity where the index of refraction is equal to:

$$n = n_0 + \frac{n_2}{2}|E|^2 = n_0 + \gamma I \tag{24}$$

where $n_0$ is the linear refraction index, $E$ is the peak electric field (derived in Eq. (17) ). $I$ is the irradiance of the laser beam within the sample, $n_2$ and $\gamma$ are related through the conversion formula $n_2(esu) = \frac{n_0 c}{40\pi}\gamma\,(m^2 / W)$ ($c(m/s)$ is speed of light).

Since our sample is thin we can approximate the Gaussian beam is parallel inside the sample. We want to calculate the phase shift of the beam when it passes through the sample. The amplitude $\sqrt{I}$ and phase of electric field in the slowly varying envelop approximation as a function of $z'$ (propagation depth in the sample), are given by two pair equation[6]:

$$\frac{d\Delta\phi}{dz'} = \Delta n(I)k \tag{25}$$

$$\frac{dI}{dz'} = -\alpha(I)I \tag{26}$$

$\alpha(I)$ contains all linear and nonlinear absorption. Using Eq.(24) we can solve the coupled Eqs. (25) and (26) together to derive $\Delta\phi$ at the exit surface of the sample as a function of stage $z$ and radial variation of incident beam.

$$\Delta\phi(z,r,t) = \Delta\phi_0(z,t)\exp(-\frac{2r^2}{w^2(z)}) \tag{27}$$

with

$$\Delta\phi_0(z,t) = \frac{\Delta\Phi_0(t)}{1+z^2/z_0{}^2}. \tag{28}$$

$\Delta\Phi_0$, the on axis ($r$ =0) phase shift at focus ($z$ =0) is defined as :

$$\Delta\Phi_0(t) = k\Delta n_0(t)L_{eff} \tag{29}$$

where $L_{eff} = (1-e^{-\alpha L})/\alpha$, $L$ is sample length and $\alpha$ linear absorption coefficient and $\Delta n_0 = \gamma I_0$ ($I_0$, $I$ at $r$=0 and $z$=0) . Now the electric field which is coming out from sample will look like

$$E_{out}(z,r,t) = E_{in}(z,r,t)e^{-\alpha L/2}e^{i\Delta\phi(z,r,t)} \tag{30}$$

where $E_{in}(z,r,t)$ is the same as in Eq.(17). Now we are going to derive the electric field in aperture. A method which is called "Gaussian decomposition" (GD) and is given by Weaire et. al.[12] can be used to obtain the far field pattern of the electric field at the aperture plane. They decompose the $E_{out}$ into a summation of Gaussian beams through a Taylor series expansion therefore from Eq. (27) and (30)

$$e^{i\Delta\phi(z,r,t)} = \sum_{m=0}^{\infty}\frac{[i\Delta\phi_0(z,t)]^m}{m!}e^{-2mr^2/w^2(z)} \tag{31}$$

therefore

$$E_a(z,r,t) = E(z,r=0,t)e^{-\alpha L/2}\sum_{m=0}^{\infty}\frac{[i\Delta\phi_0(z,t)]^m}{m!}\frac{w_{m0}}{w_m}\times\exp(-\frac{r^2}{w_m{}^2}-\frac{ikr^2}{2R_m}+i\theta_m). \tag{32}$$

defining $d$ as distance from sample to aperture and $g = 1 + d/R(z)$ (R(z) is defined in Eq.(15) ) all parameters in Eq.(32) are expressed as[13]:

$$w_{m0}^2 = \frac{w^2(z)}{2m+1}, \quad d_m = \frac{kw_{m0}^2}{2}, \quad w_m^2 = w_{m0}^2[g^2 + \frac{d^2}{d_m^2}]$$

$$R_m = d[1 - \frac{g}{g^2 + d^2 / d_m^2}]^{-1} \text{ and } \theta_m = \tan^{-1}[\frac{d / d_m}{g}] \tag{33}$$

The GD method is very useful for small phase distortions detected with Z-scan therefore only few terms of Eq.(32) are needed. Now we can calculate transmittance power through the aperture:

$$P_T(\Delta\Phi_0(t)) = c\varepsilon_0\pi\int_0^{r_a} |E_a(r,t)|^2 rdr \tag{34}$$

Including the pulse temporal variation, the normalized Z-scan transmittance can be calculated as

$$T(z) = \frac{\int_{-\infty}^{\infty} P_T(\Delta\Phi_0(t))dt}{S\int_{-\infty}^{\infty} P_i(t)dt} \tag{35}$$

where $P_i(t) = \pi w_0^2 I_0(t)/2$ is the instantaneous input power (within the sample) and $S = 1 - \exp(-2r_a^2 / w_a^2)$ is the aperture linear transmittance ($w_a$ is the beam radius at the aperture). In above discussion we have assumed the effect of third order nonlinearity only and that no absorptive nonlinearity effects that arise from multiphoton or saturation absorption exist. Multiphoton absorption suppress the peak and enhance the valley, while saturation produce the opposite effect[6,9].

### c. The Dual-Beam technique

Shen et al.[12-14] introduced a pump probe laser scheme under mode-mismatched and mode-matched conditions to make sensitive TL measurements. In such dual beams experiments, one of laser beams is essentially probing the effect of the TL caused by the pump beam by scanning across its focus. This results in an effective z-scan of the probe beam across a TL generated by a focusing pump beam. The closed aperture case of this scenario is shown schematically in Fig.5 which has been used by Marcano et. al.[15] to measure the absorption coefficient of water with high accuracy.

Mathematically, as in Ref. [15], we can also use the expression of Shen et. al.[12-14], who have derived an expression for the TL signal using diffraction approximation for Gaussian beams in steady state case as:

$$S(z) = [1 - \frac{\theta}{2}\tan^{-1}(\frac{2mV}{1+2m+V^2})]^2 - 1 \tag{36}$$

where

$$m = (\omega_p / \omega_o)^2,$$

$$V = (z - a_p)/z_p + [(z - a_p)^2 + {z_p}^2]/[z_p(L - z)],$$

$$\omega_{p,o} = b_{p,o}[1 + (z - a_{p,o})^2 / z^2{}_{p,o}]^{1/2}, \tag{37}$$

$$\theta = -P_o \alpha_0 l (ds/dT)/\kappa\lambda_p$$

$z$ is sample position with respect to the focal point, $a_p$, $a_o$, $z_p$, $z_o$ and $b_p$, $b_o$, are position of the waists, the confocal parameters and the beam radius for the probe and pump beams, respectively. $\lambda_p$ is the wavelength of the probe beam, $\kappa$ is thermal conductivity coefficient of the sample. $L$ is the detector position, $P_o$ is the total power of the pump beam and $l$ is the sample thickness.

Both continuous and pulsed lasers have been effectively used for z-scan experiments that have relied on these mathematical principles discussed here[19]. These discussions in this section form the basis of our present work that we present hereafter. We explore the open-aperture dual beam TL experiments and achieve our single experimental setup to achieve high sensitive measurements.

## III. Present Work

Our experiments are variation from the conventional z-scan discussed in the above section. We not only use the single beam technique as mentioned in the previous section, but with very simple changes in the experimental set-up, make measurements corresponding to the dual beam z-scan experiments. We will now concentrate more on our actual experimental scheme and the results and discussions arising thereafter.

### a. Experiment

Our experimental scheme involves a sub-100 femtosecond mode-locked Er:doped fiber laser (IMRA Inc.) operating at a repetition rate of 50MHz and provides the fundamental (1560nm) and its second-harmonic wavelength (780nm) simultaneously as a single output. The pulse characteristics of the laser pulses are shown in Fig.6. Either we use both the wavelengths from the laser simultaneously or separate the two copropagating beams with the help of a dichroic beamsplitter and use each of them independently. We scan the sample through the focal point of a 75cm focusing lens and this allows a smooth intensity scan for either/or both of the wavelengths. Care has been taken to make sure that there is no effect of the laser beams on the cuvette alone by conducting an empty cuvette experiment. A silicon photodetector (Thorlab: DET210) is used for the 780nm beam detection, while an InGaAs photodetector (Acton Research) is used for the 1560nm beam detection.

We find that the 1560nm beam produces changes in the relative transmission of the laser beam at different intensities as the sample is scanned through the lens focus depending whether we collect all the light or only central 40% of the transmitted light (Fig.7). These results in Fig.7 essentially represent the z-scan technique of Sheik Bahaei et al.[6] to measure the real and imaginary parts of the third-order optical nonlinearity ( $\chi^{(3)}$ ). However, the 780nm beam does not produce any effect even at our peak powers at the focal point of the laser as is expected from negligible absorption at 780nm (Fig. 7c). This enables us to use the 780nm wavelength as the non-interacting probe beam for the subsequent dual-beam experiments where we use both the wavelengths from the laser simultaneously. Since our 75cm lens focuses the 780nm probe beam to its minimal spot size position 0.4mm ahead of the pump beam of 1560nm, this is a mode-mismatched pump-probe experiment. However, the focal spot size of 9µm for 780nm is 15µm smaller than the corresponding 1560nm spot size at its own focus and from Eq.(24) the Rayleigh range for 780nm is 0.32mm and for 1560mm it is 1.15mm. Thus, the 780nm laser volume is always confined within the 1560nm laser beam volume when both the beams are used simultaneously from the laser and can act as an effective probe.

### b. Results and Discussion

Fig.8 shows the results of the experiment when we collect the 780nm probe beam only by separating the 1560nm pump beam after passing through the sample at different intensities as the sample is scanned through the lens focus depending whether we collect all the light or only central 40% of the transmitted light. Fig.8a shows the case when the entire transmitted probe (780nm) beam is being collected and this essentially depicts the saturation environment created by the pump (1560nm) beam. This statement is further reinforced by Fig.8b, where the case of 1560nm beam alone from Fig.7a is plotted along with the results in Fig.8a. Essentially, as the 1560nm beam starts to saturate the sample at its focal point, the 780nm beam also experiences a saturated environment, whereby its transmittance increases at its focal point and shows an identical transmission behavior although the signal level is two orders of magnitude lower. Such a result indicates the thermal capacity of water that can affect the spectroscopic behavior of water. Finally, Fig.8c represents the thermal lens effect of the pump beam resulting in a temperature gradient, which in turn produces a spatial gradient in refractive index which is depicted in the relative change in transmittance of the probe beam. Such thermal lensing (TL) effect can be used to determine the absorption coefficient ($\alpha_0$) of the sample at the pump wavelength very accurately[15].

The solid line in the fig.8c is the result of a theoretical fit to Eq.(36). This fit gives the value of phase shift, $\theta$ = 9.957, which when substituted in Eq.(37) with the parameters $ds/dT = -9.1\times10^{-5}\ K^{-1}$ and $\kappa = 0.598\times10^{-2}\,WK/cm$ for pure water[20], we get the calculated value for $\alpha_0$ as 10.6327cm$^{-1}$ for the 1560nm beam which is within 1% of reported literature data[21]. While this fairly large value of $\alpha_0$ need not be measured with such a sensitive technique, it serves as the proof-of-principle for the experimental setup and the accuracy of the experimental measurements.

Our experimental results discussed above also enables us to determine the nonlinear absorption coefficients of water. For nonlinear materials the index of refraction $n$ is expressed in terms of nonlinear $n_2$ or $\gamma$ through the relation: $n = n_0 + \frac{n_2}{2}|E|^2 = n_0 + \gamma I$, where $n_0$ is the linear index of refraction, $E$ is the peak electric field (cgs) and $I$ (MKS) is the intensity of the laser beam inside the sample. $n_2$ and $\gamma$ are related to each other as: $n_2(esu) = \frac{cn_0}{40\pi}\gamma(m^2/W)$, where $c$(m/s) is the speed of light in vacuum. The third order optical susceptibility is considered to be a complex quantity: $\chi^{(3)} = \chi_R{}^{(3)} + \chi_I{}^{(3)}$. The real and imaginary parts are related to the $\gamma$ and $\beta$ respectively[22] where $\beta$ is the nonlinear absorption coefficient and is defined as $\alpha(I) = \alpha_0 + \beta I$. We fit the fully open aperture data with 1560nm wavelength alone from Fig.5a (solid line in Fig. 5a) to a theoretical expression derived by solving the differential equation[23] for the transmitted light through a sample of thickness $l$

$$T(z) = \eta + \frac{\beta I_0 l}{(1 + z^2 / z^2{}_0)} \tag{38}$$

where $z_0 = kw_0{}^2/2$, $k$ and $w_0$ are the wave vector and the minimum spot size in the focal point respectively, while $\eta$ and $\beta$ are the fitting parameters. The best fit gives the value of $\beta$ = -2.58 cm/GW.

In Fig.5b, the valley-peak structure representing the 40% closed-aperture data for 1560nm suggests a self-focusing effect inside the sample. The Ryleigh range (Z(r)) for 1560mm is 1.15mm. From fig. 5b, valley to peak separation at 1560nm is 4mm, which is 3.4×Z(r) indicating that all the effects at 1560nm are thermal in nature[24]. So we use the Gaussian Decomposition method to fit this closed aperture z-scan data quite convincingly (fig. 5b, solid line fit to the raw data), and we obtain $\gamma=1.57\times 10^{-3}$ cm$^2$ /GW, which is proportional to $n_2=4.9\times10^{-12}$ esu. Thus $\alpha_0$, $\beta$ and $n_2$ values of the water sample at 1560nm wavelength are determined.

Finally, we use the theoretical expression for the thermal lens of the pump beam given by Eq.(38) to fit the experimental data (Fig.6a) of the probe beam in case of the dual beam experiment. The solid line in fig.6a is the theoretical fit and results in a calculated parameter $\beta = -8.5\times10^{-3}$ cm/GW, which is extremely small and indicates the high sensitivity of our experiments. This indicates the very small induced effect in the probe beam of 780nm by the presence of the pump beam of 1560nm causing thermal lens. Thus, this technique measures the nonlinear absorption coefficient of materials at wavelengths with negligible linear absorption.

## IV. Acknowledgements

The authors thank the Ministry of Information Technology, Govt. of India, for partial funding for the research results presented here.

Figure 1. Schematic of a Gaussian beam of waist ($w_{01}$) propagating through a thin lens of focal length $f$. The beam focuses at distance $l$ from the lens with a Rayleigh range of $RR$.

Figure 2. Measurement of laser beam size by translating a knife-edge across the laser beam falling on a photodiode.
(a) A plot of the intensity of the light (along y-axis) coming to the photodiode as the translating knife-edge (distance along x-axis) lets out more and more of the incident laser beam into the photodiode.
(b) A derivative of the data in Fig. 2a gives the beam size in the lower plot, which fits to a Gaussian beam waist of 2.77mm.

Figure 3. Schematic of typical z-scan experimental setup where a sample (S) is scanned across a laser beam that is being focused through a lens of focal length (L1) and is collected through an aperture (A) and a lens of focal length (L2) into the detector (D).

Figure 4. Typical experimental results of z-scan experiments performed under various conditions. (a) For materials with positive refractive index with 40% closed aperture in front of the detector, the typical increased signal followed by reduced signal at the detector is due to the characteristic beam bending as shown. (b) For materials with negative refractive index with 40% closed aperture in front of the detector, the results are opposite to that of 4a due to the opposite bending of the beam. (c) Experiments with open aperture result in a characteristic dip or a peak at the focal point position of the scan either due to multiphoton absorption or absorption saturation.

Figure 5. Typical experimental results of dual-beam thermal lens experiment where the probe beam is collected through a 40% closed aperture.

Figure 6. The dual wavelength femtosecond fiber laser (IMRA Inc., Femtolite-C) pulse characteristics as measured in our laboratory. (a) The spectra at the center wavelength of 780nm measured through the SP-150 monochromator (Acton Research Co.) into a silicon photodetector (Thorlab: DET210). (b) The pulsewidth of the 780nm pulse measured through a non-collinear autocorrelation using a speaker as the delay arm into a second-harmonic BBO crystal, which is detected into a PMT (Hamamatsu 1P28). The Gaussian fit to the autocorrelation trace provides a pulsewith of 90fs for the 780nm laser pulses.
(c) The spectra at the center wavelength of 1560nm measured through the SP-150 monochromator into an InGaS detector (both Acton Research Co.) (d) The pulsewidth of the 1560nm pulse measured through collinear cross-correlating with the 780nm beam into a second-harmonic BBO crystal, which is detected into a PMT (Hamamatsu 1P28). Using a Gaussian fit and deconvoluting the 90fs pulse of 780nm results in a pulsewidth of 120fs for the 1560nm pulse.

Figure 7. (a) Measured z-scan of a 16mm thick double distilled water using 95fs pulses at $\lambda$=1560nm (diamond) and theoretical fit (solid line) for fully open aperture (experimentally measured transmittance is normalized to unity).

(b) Measured z-scan of a 16mm thick double distilled water using 95fs pulses at $\lambda$=1560nm (diamond) and theoretical fit (solid line) for 40% closed aperture (experimentally measured transmittance is normalized to unity).
(c) Water spectra covering the 780nm and 1560nm range of wavelengths.

Figure 8. (a) Measured z-scan transmittance of 80fs pulses of $\lambda$=780nm as a probe through a 16mm thick double distilled water being irradiated with 95fs of $\lambda$=1560nm as pump (diamond) and theoretical fit (solid line) in fully open aperture (raw transmittance data is presented to illustrate sensitivity of our measurements).
(b) Replot of Fig. 5a of 1560nm alone and Fig. 6a of 780nm probe measurements for 1560nm pump case to show that both essentially have same features except their two orders of magnitude difference in their signal levels.
(c) Measured z-scan transmittance of 80fs pulses of $\lambda$=780nm as a probe through a 16mm thick double distilled water being irradiated with 95fs of $\lambda$=1560nm as pump (diamond) and theoretical fit (solid line) in 40% closed aperture (experimentally measured transmittance is normalized to unity).

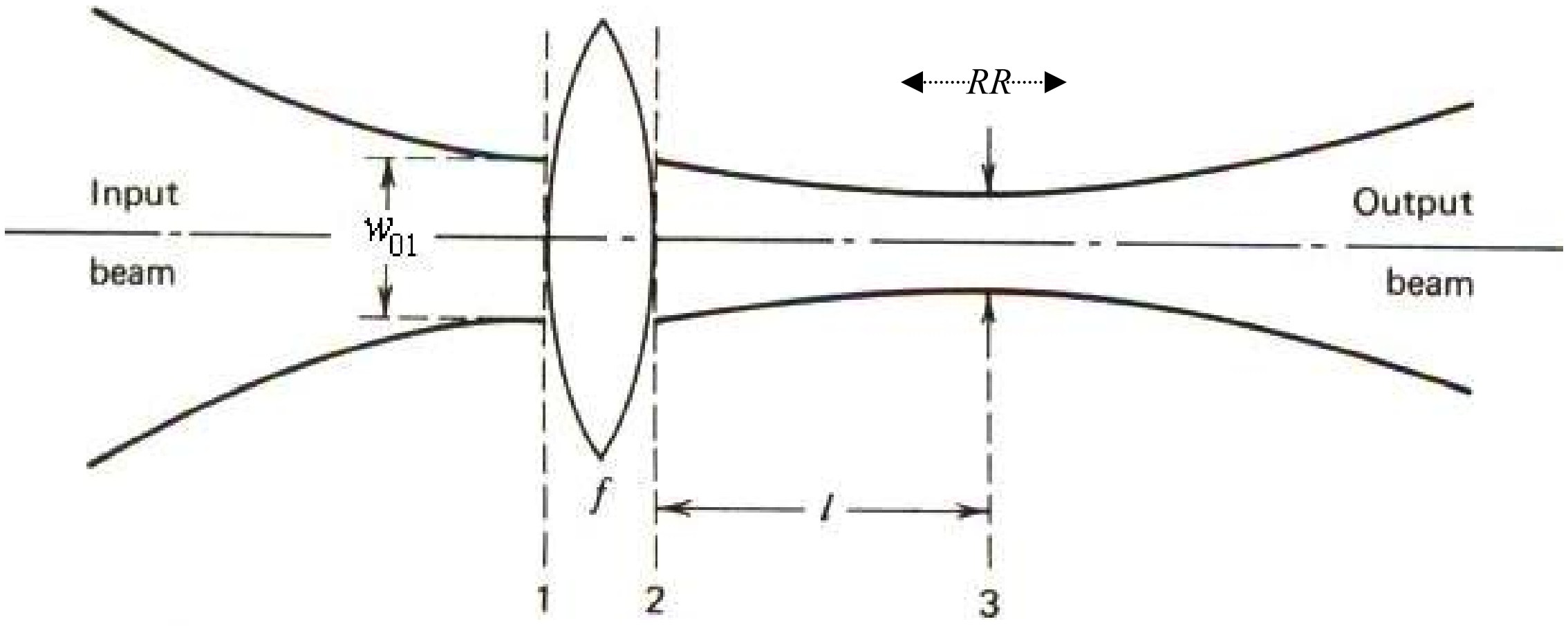

Figure 1

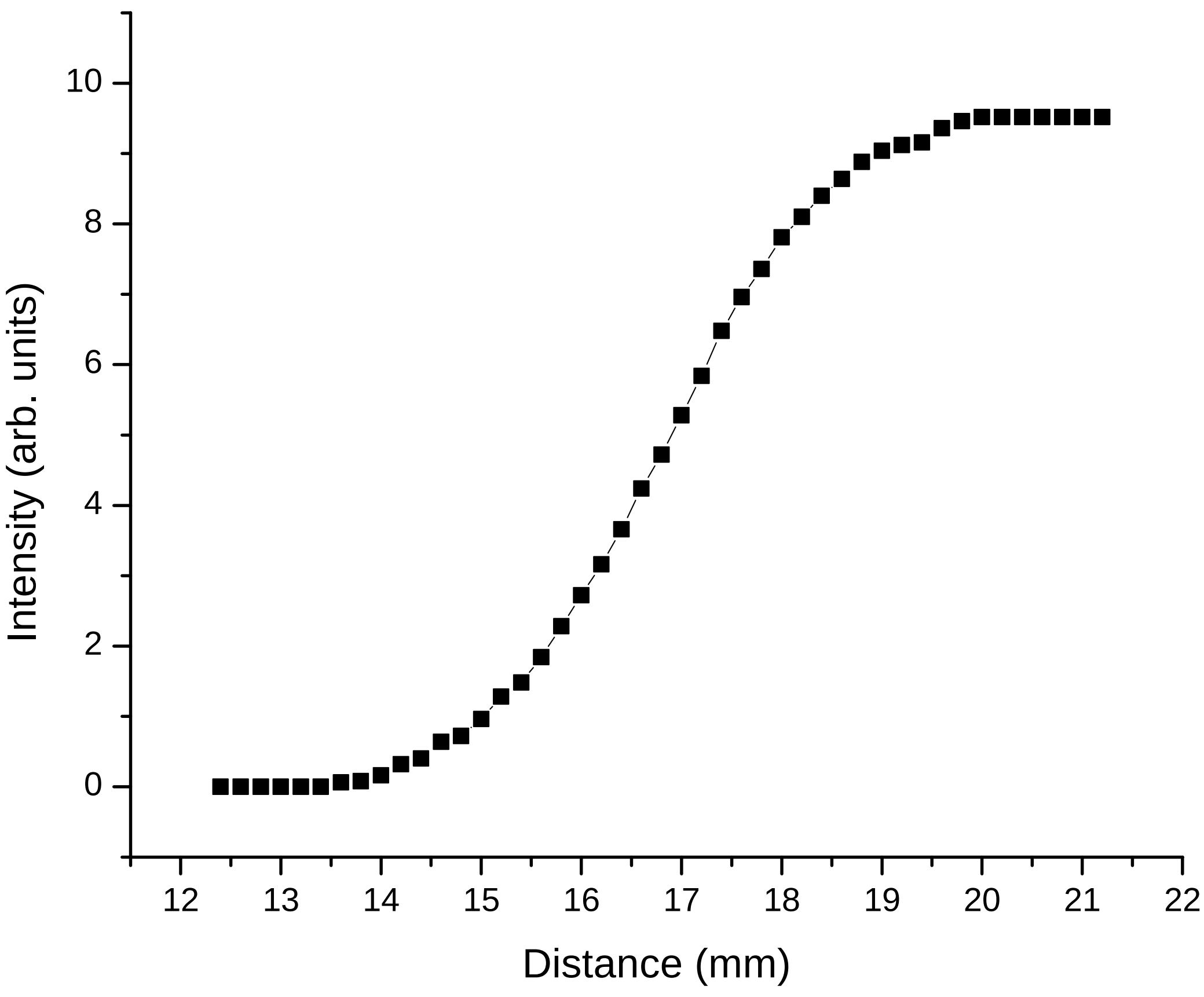

Figure 2a

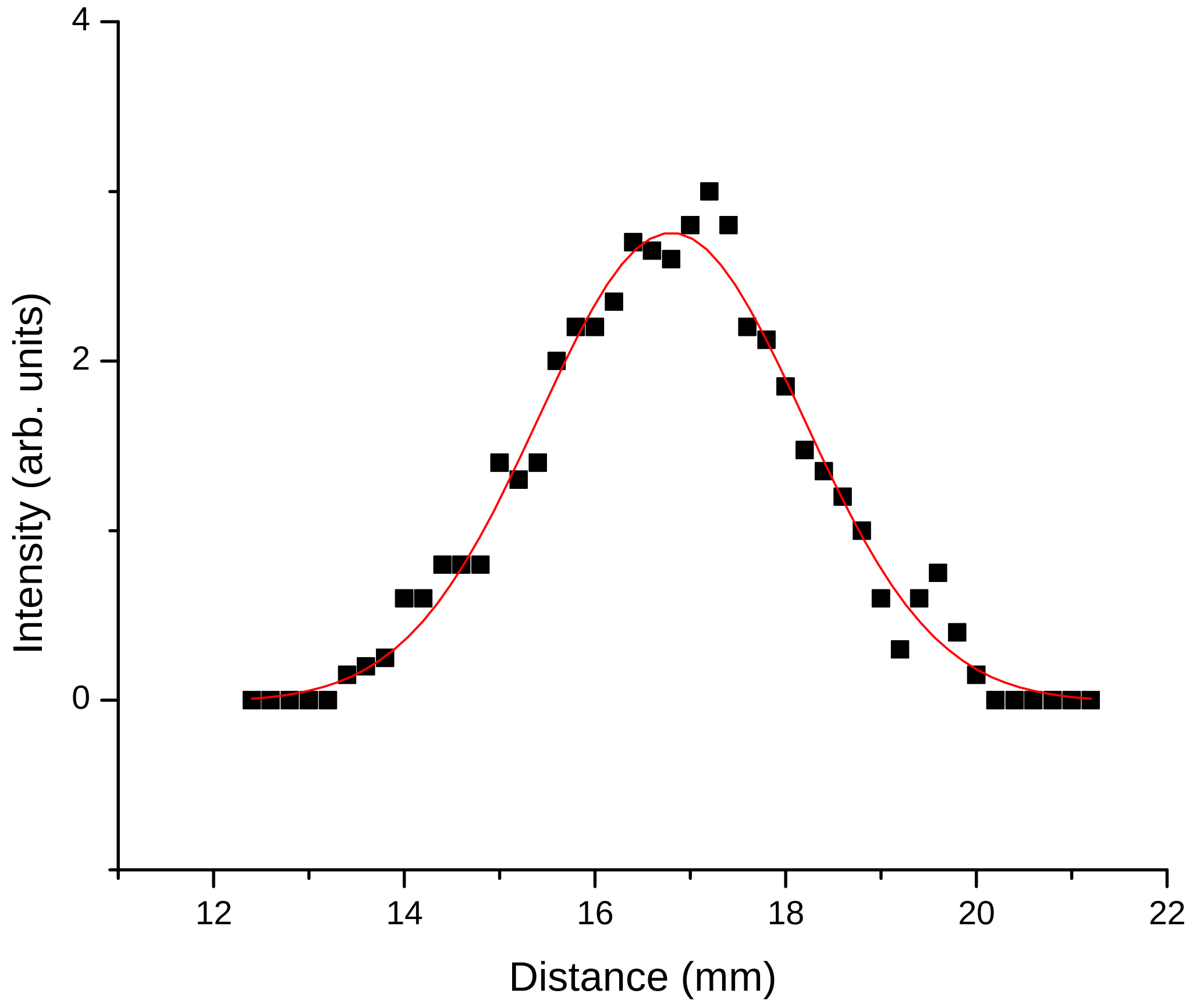

Figure 2b

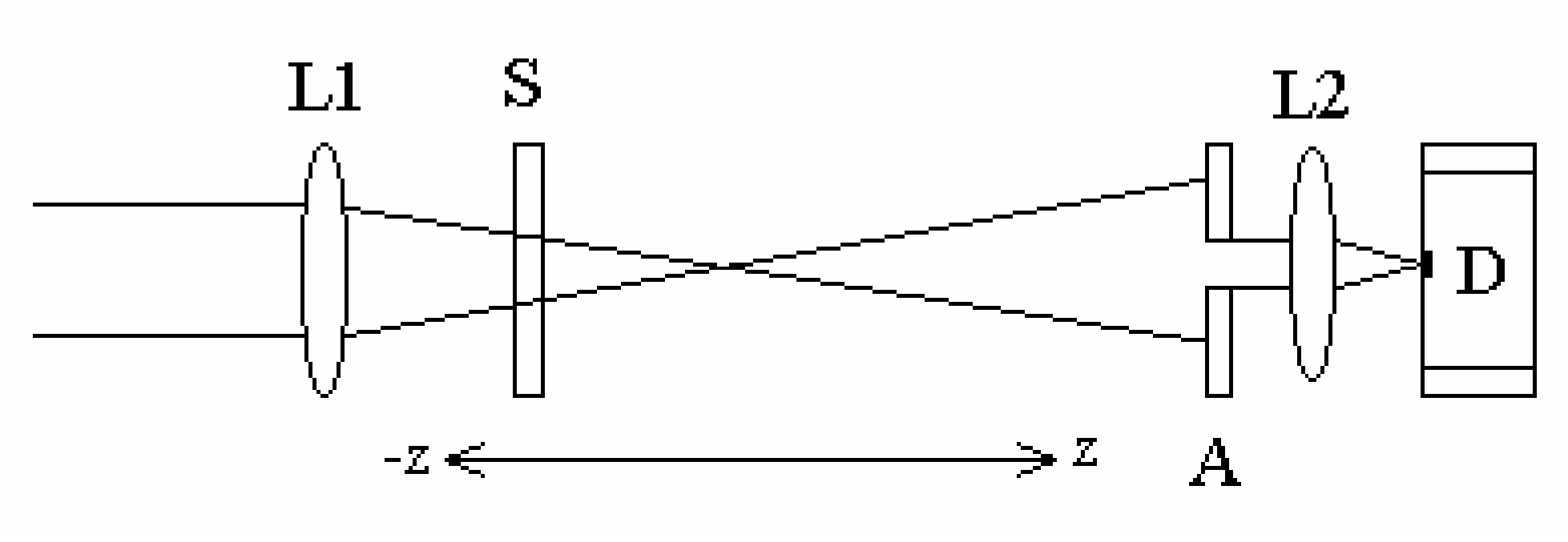

Figure 3

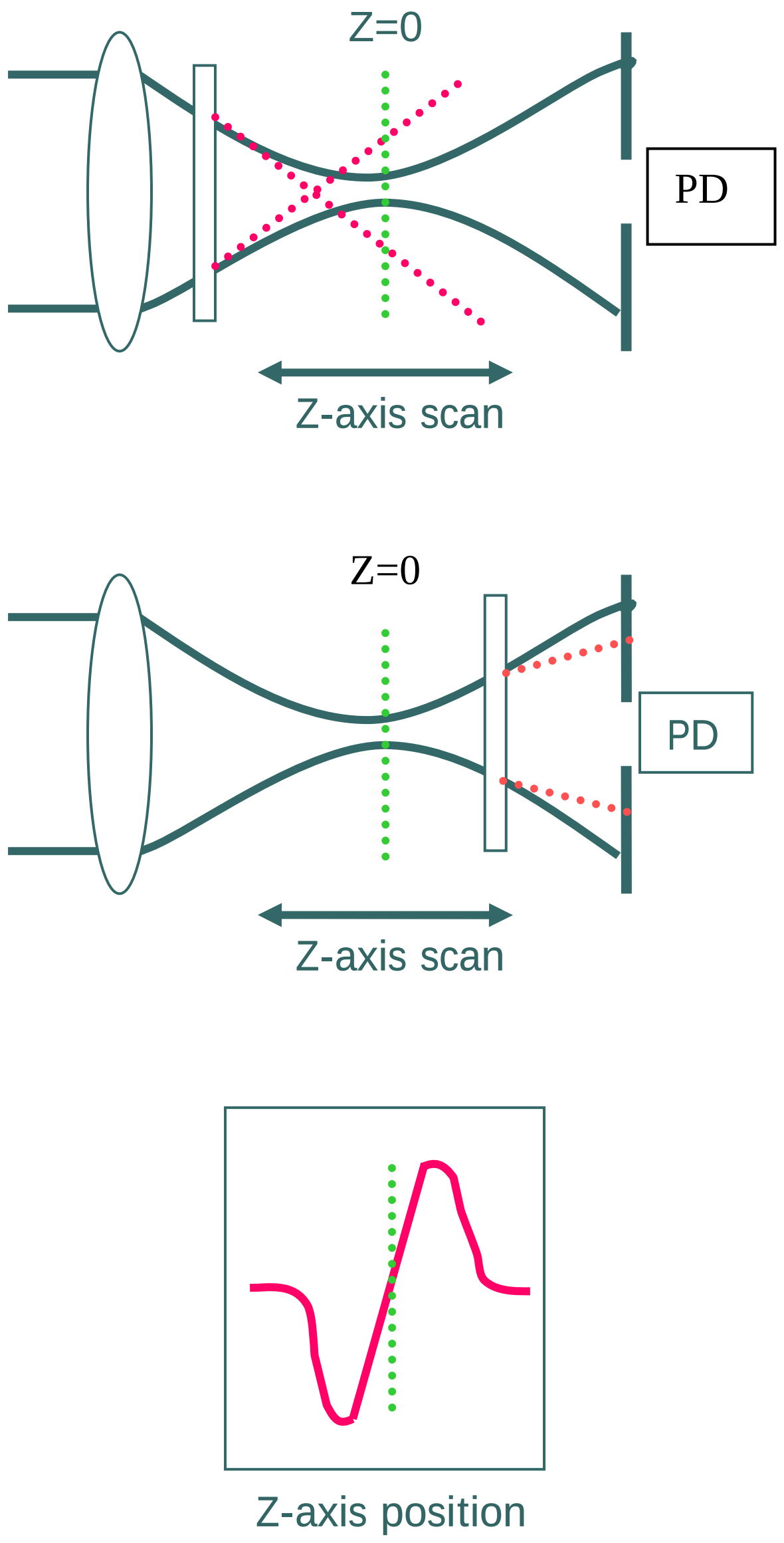

Figure 4a

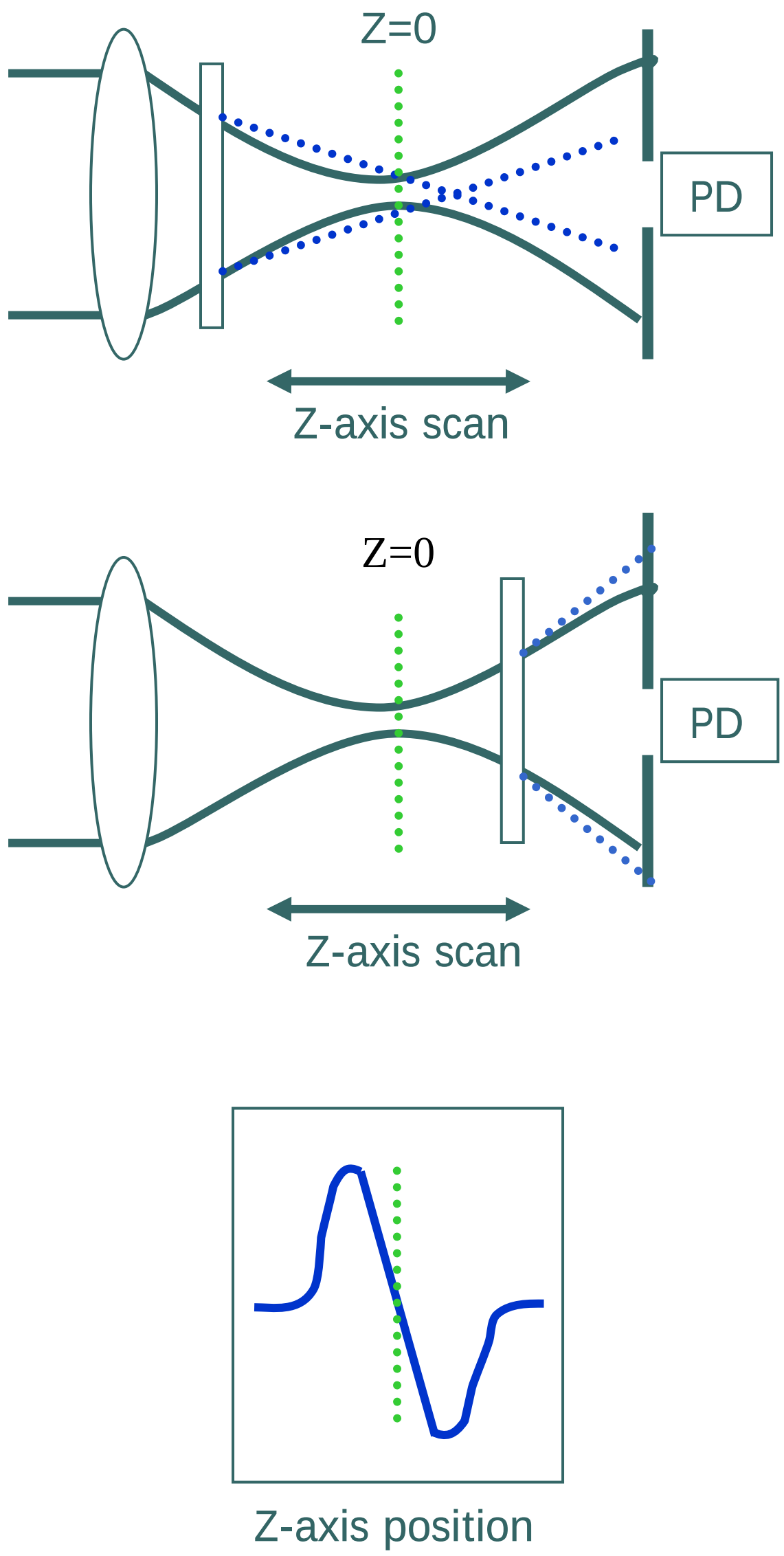

Figure 4b

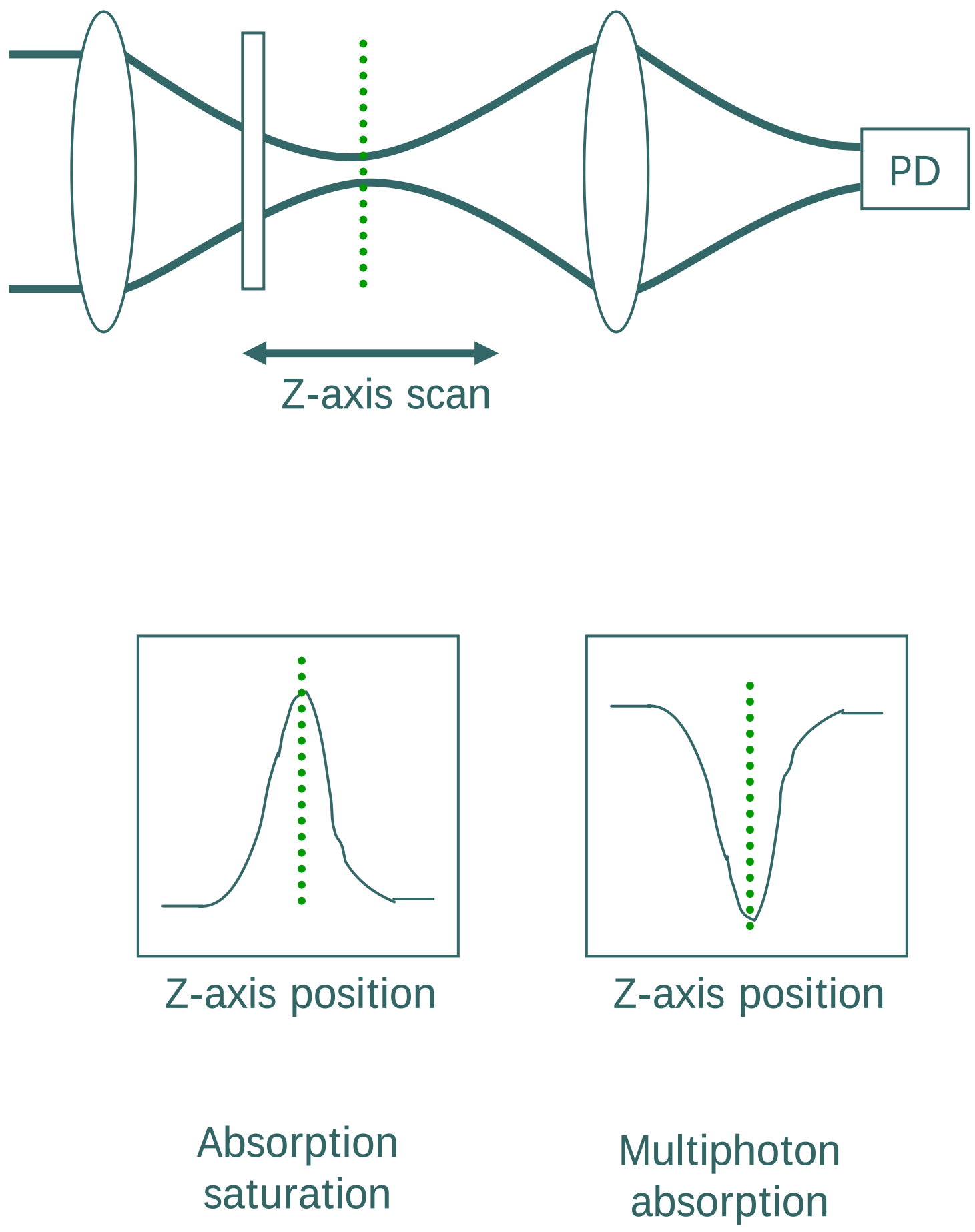

Figure 4c

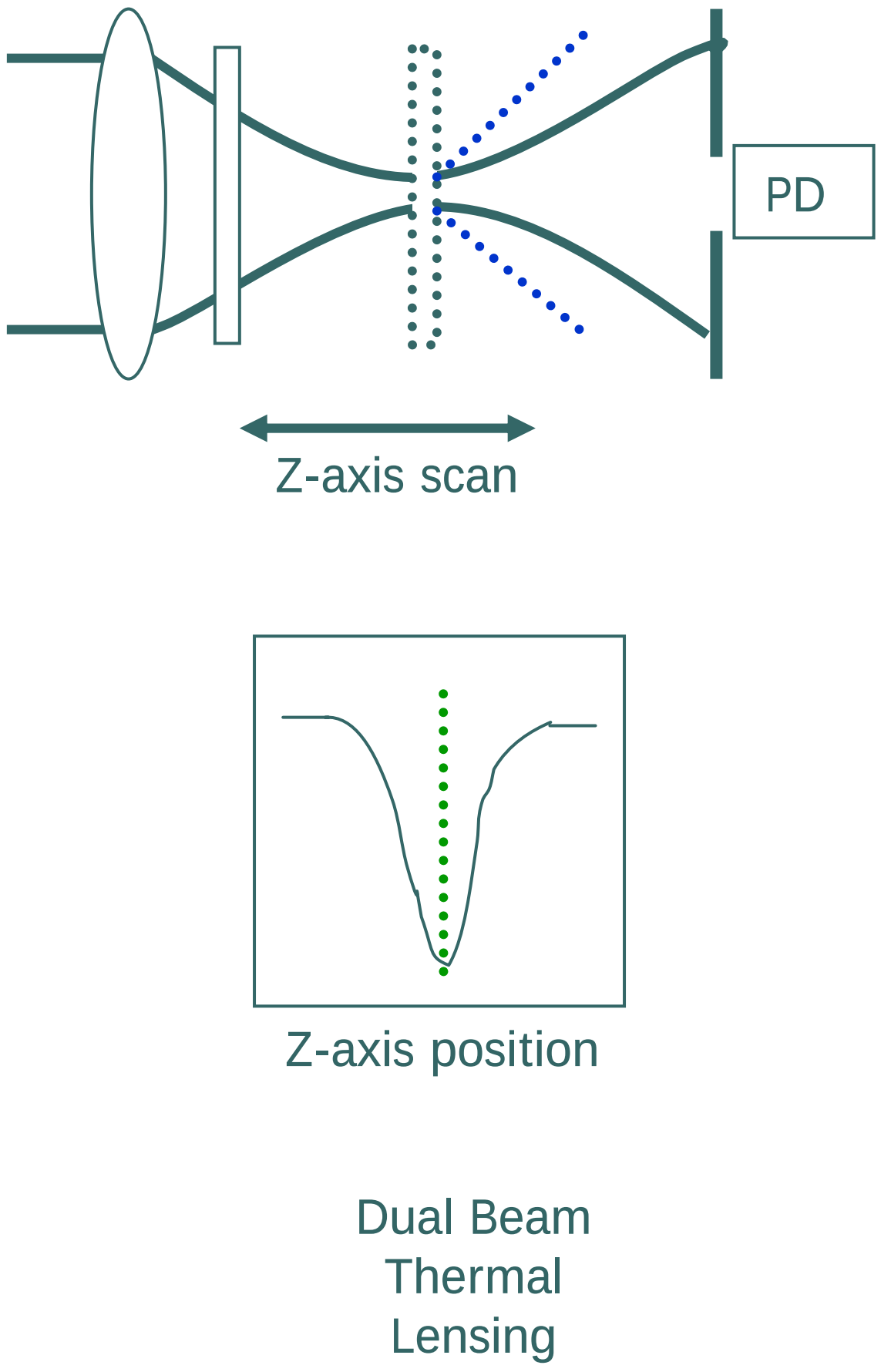

Figure 5

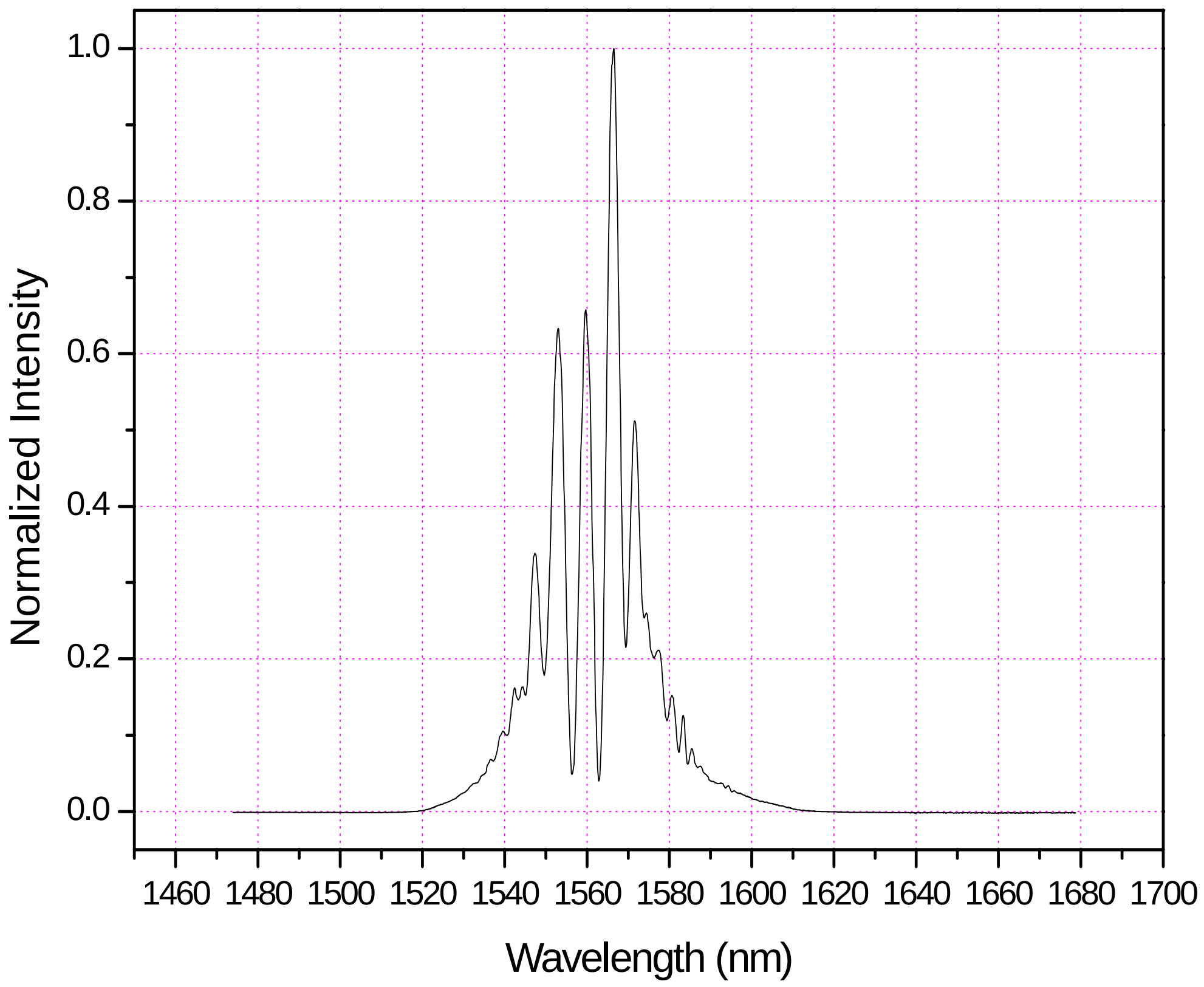

Figure 6c

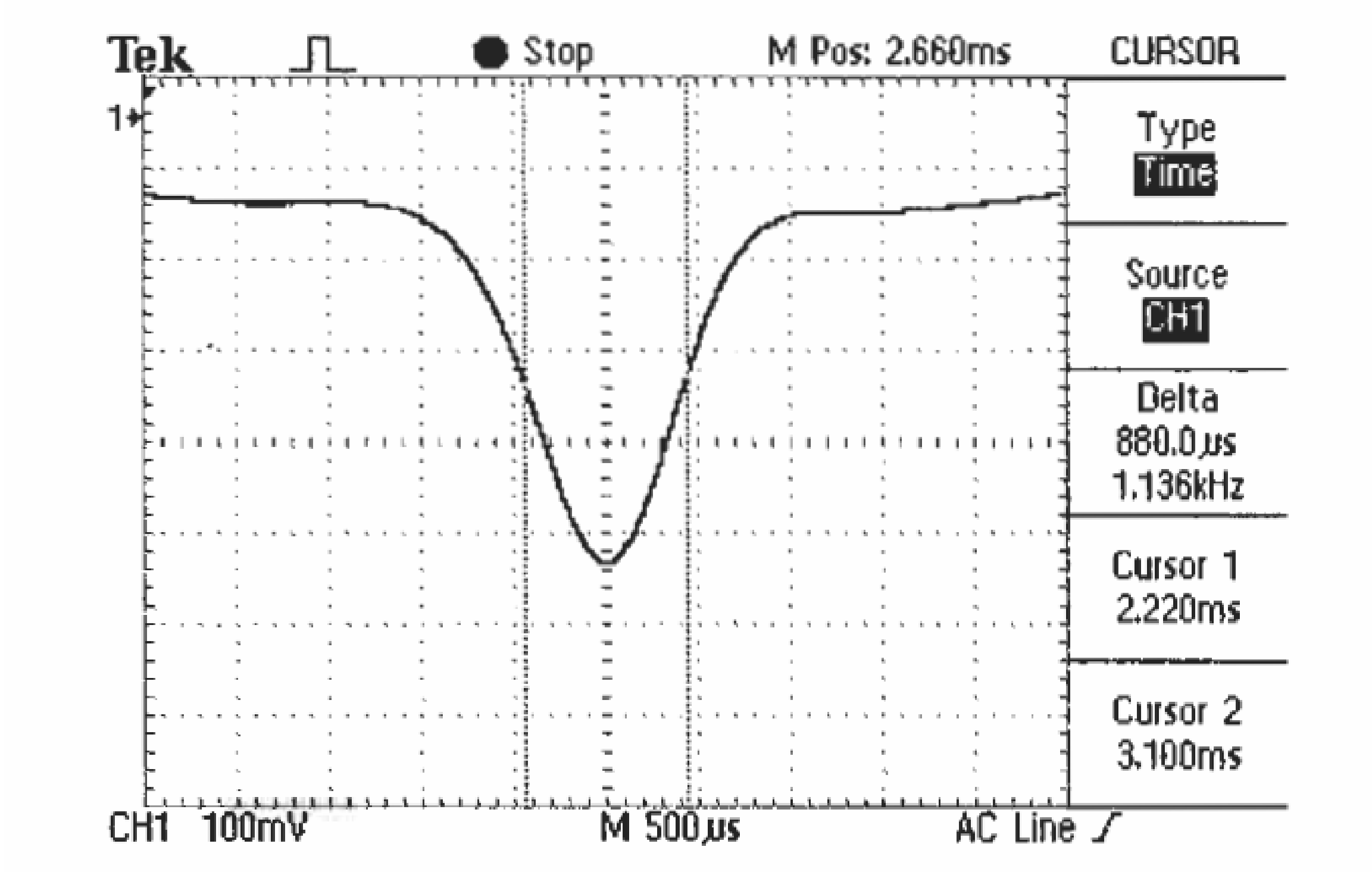

Figure 6d

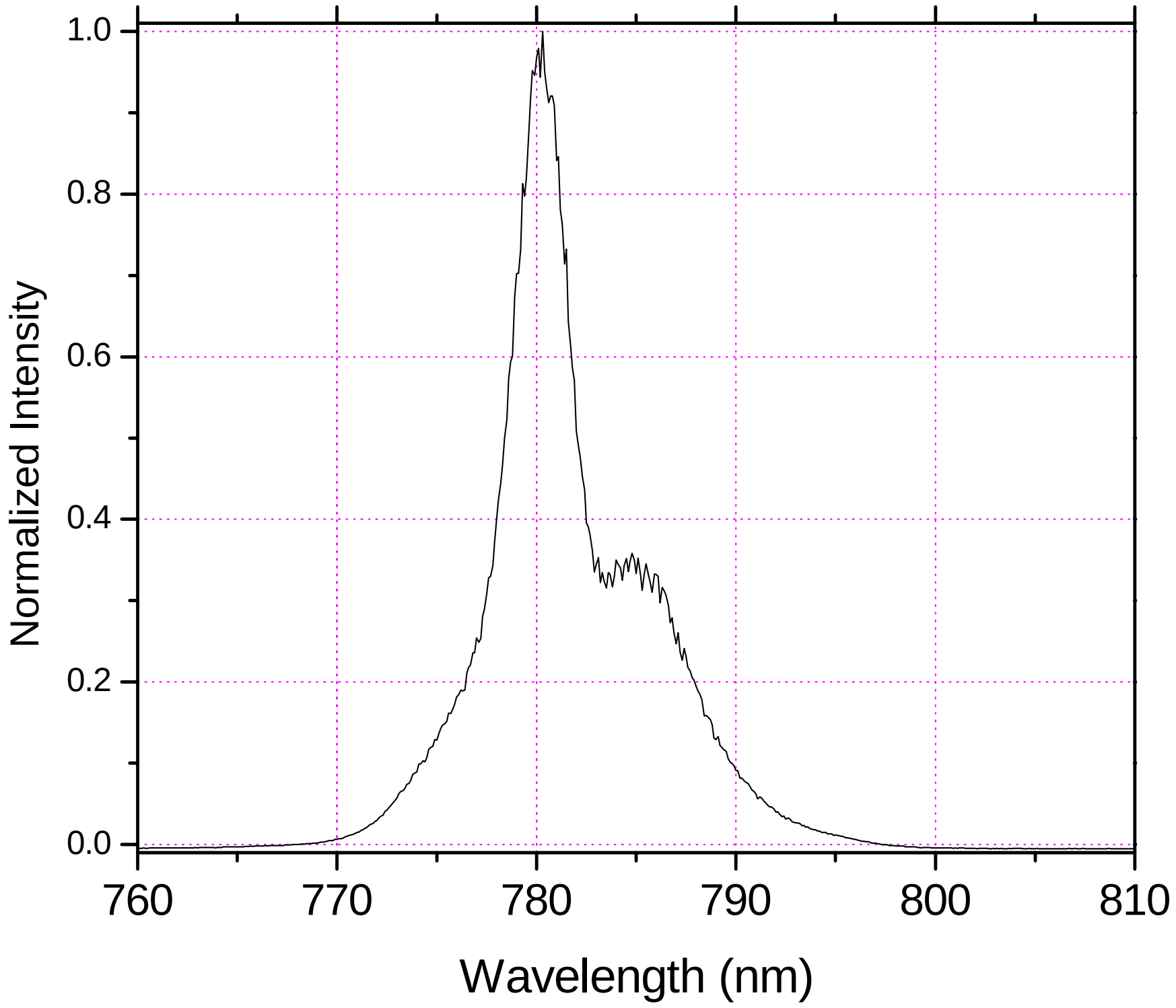

Figure 6a

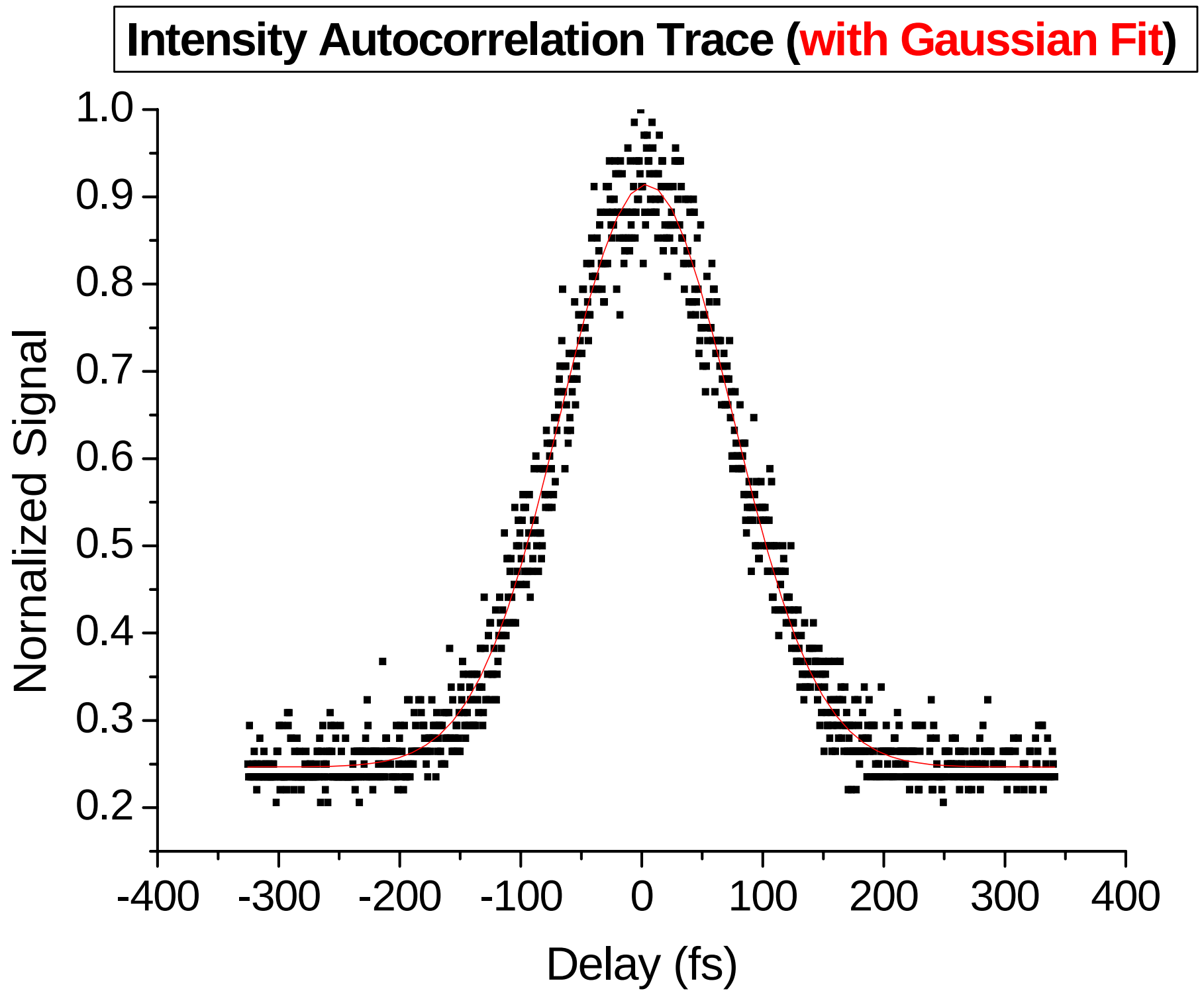

Figure 6b

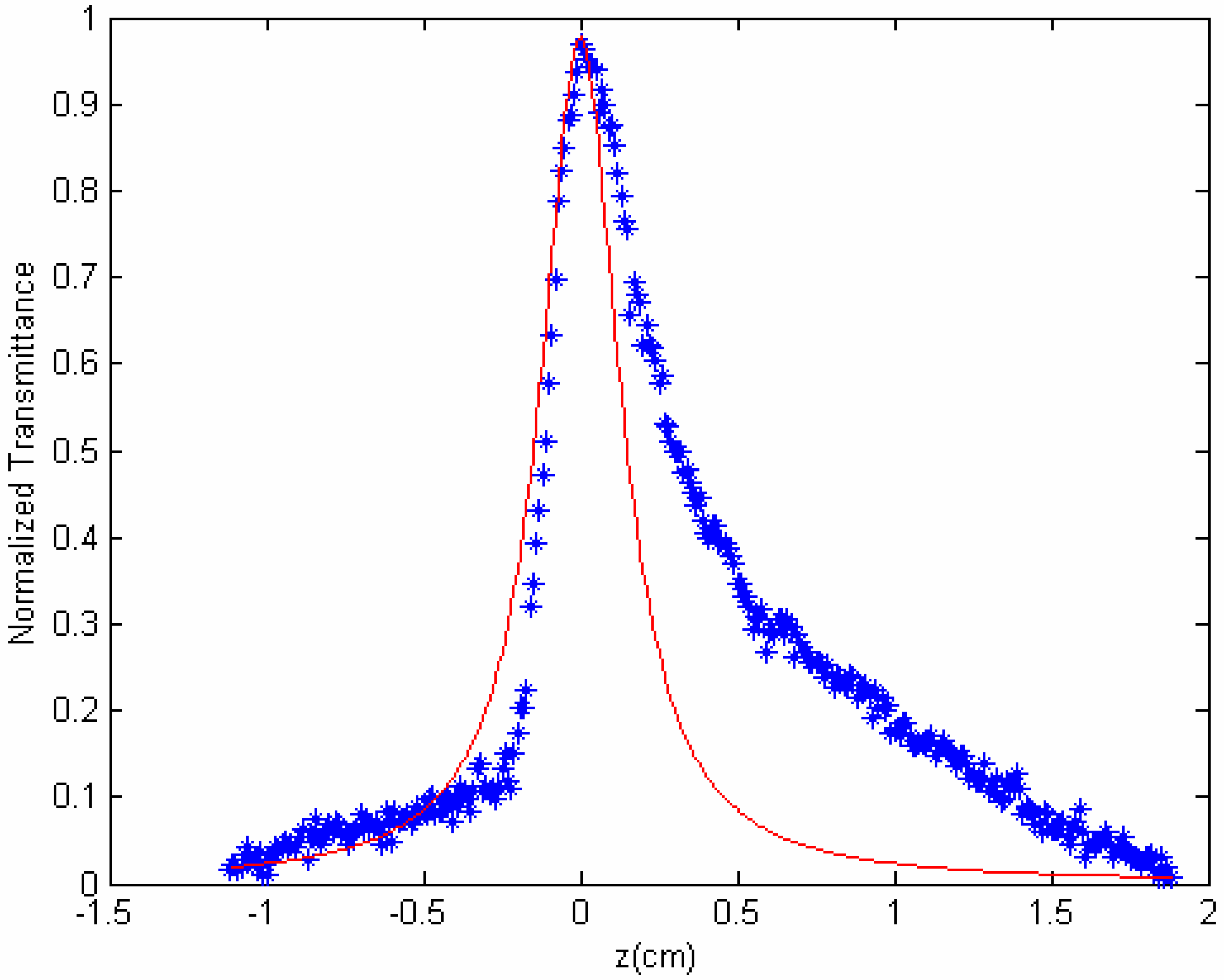

Figure 7a

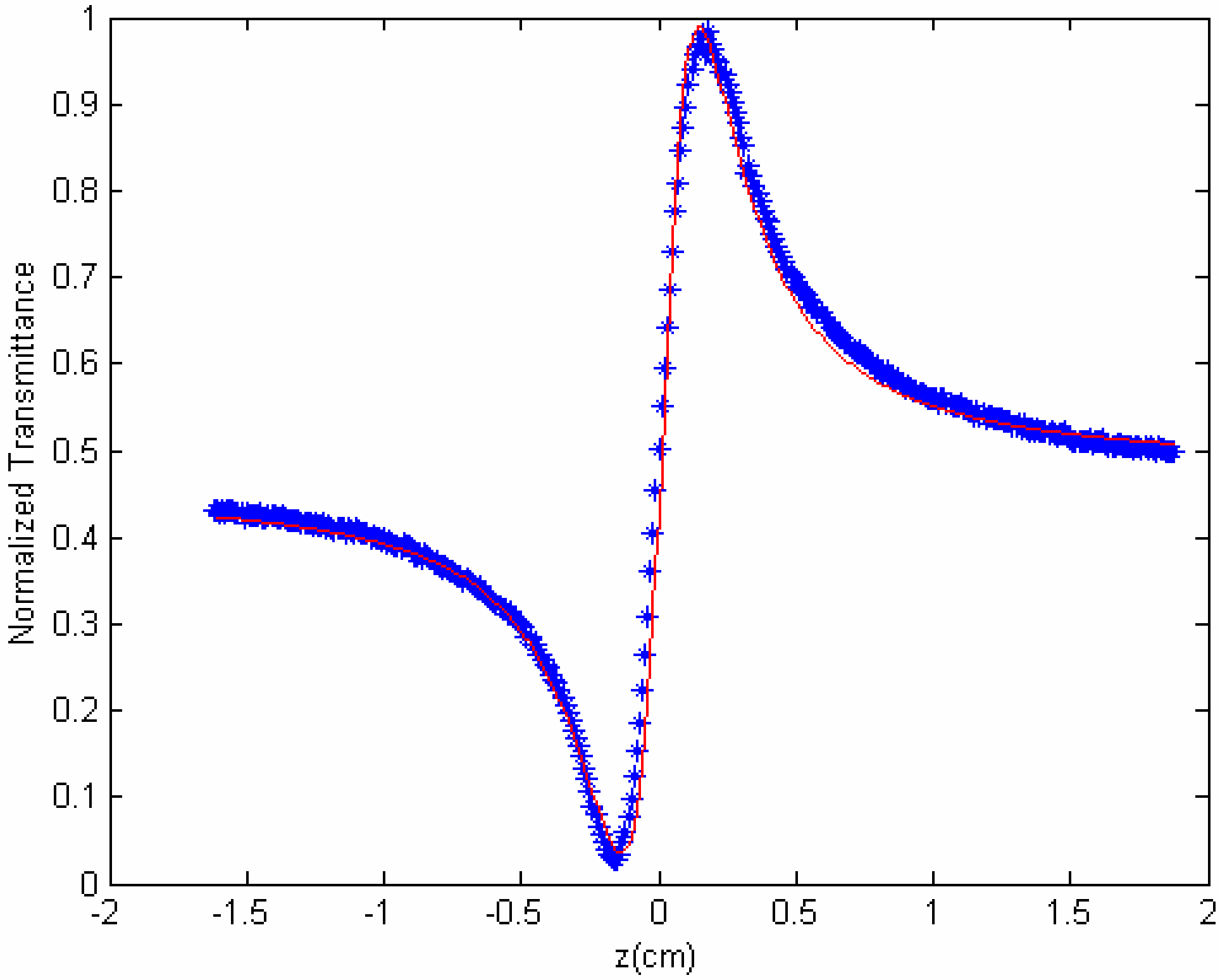

Figure 7b

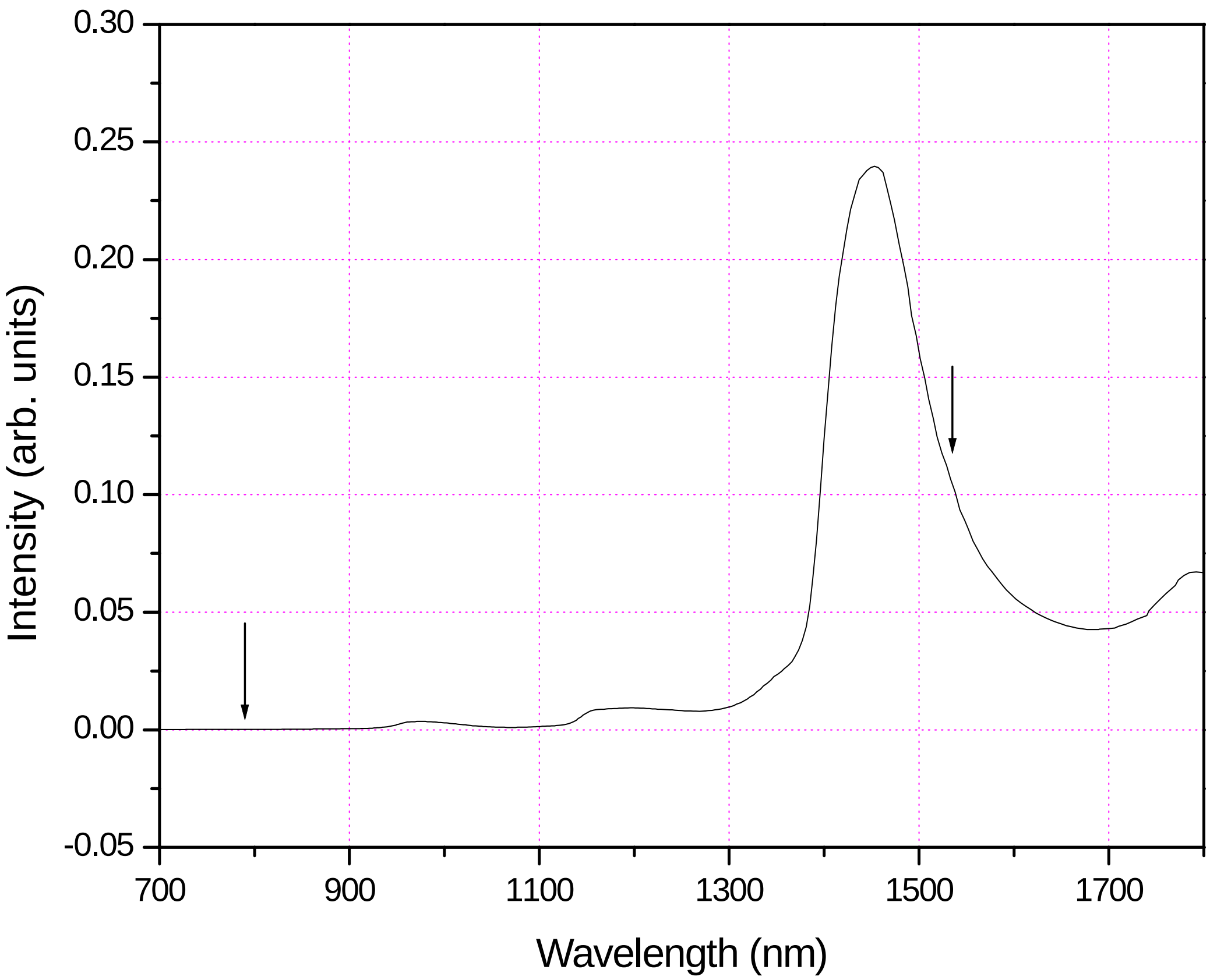

Figure 7c

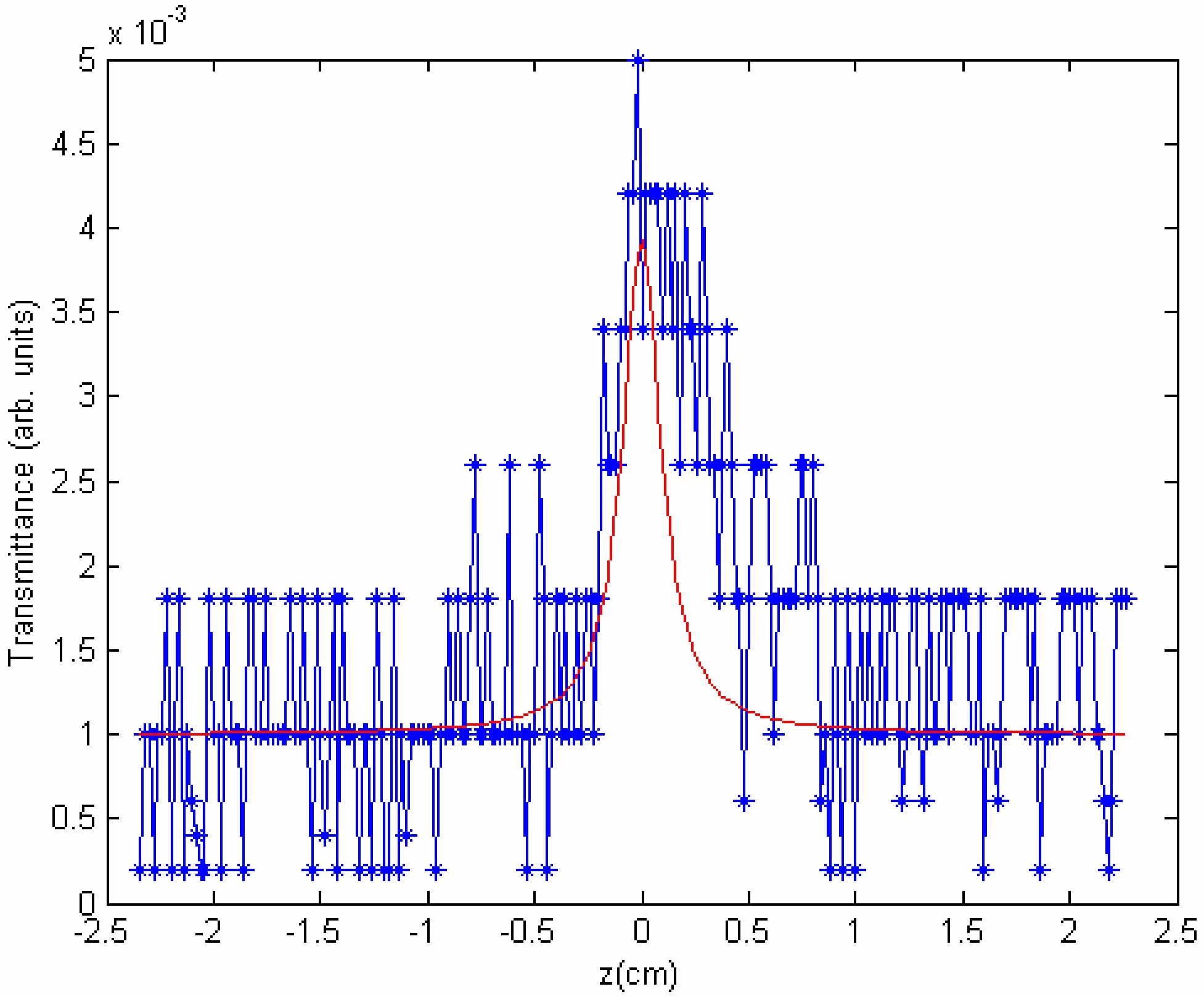

Figure 8a

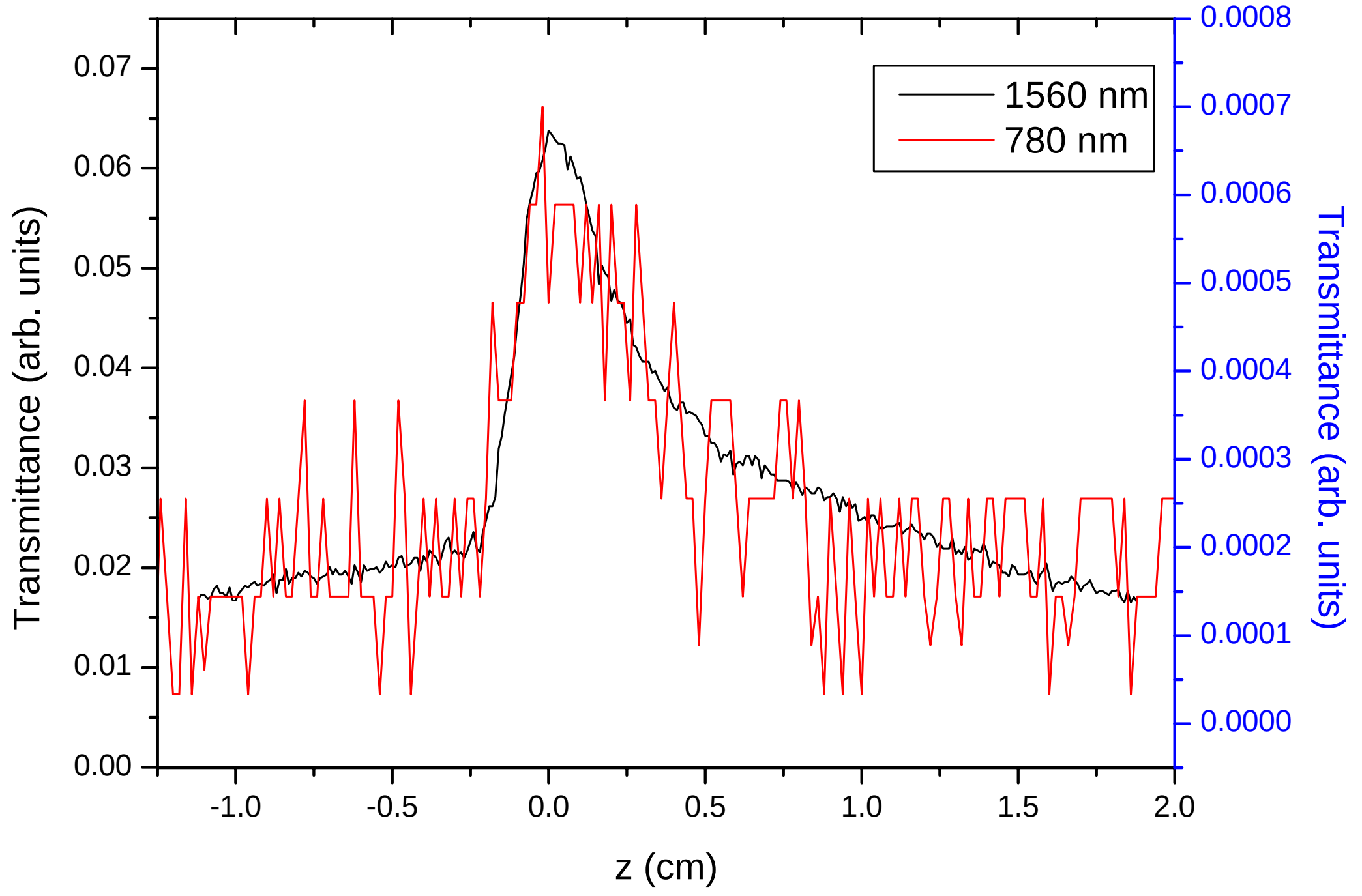

Figure 8b

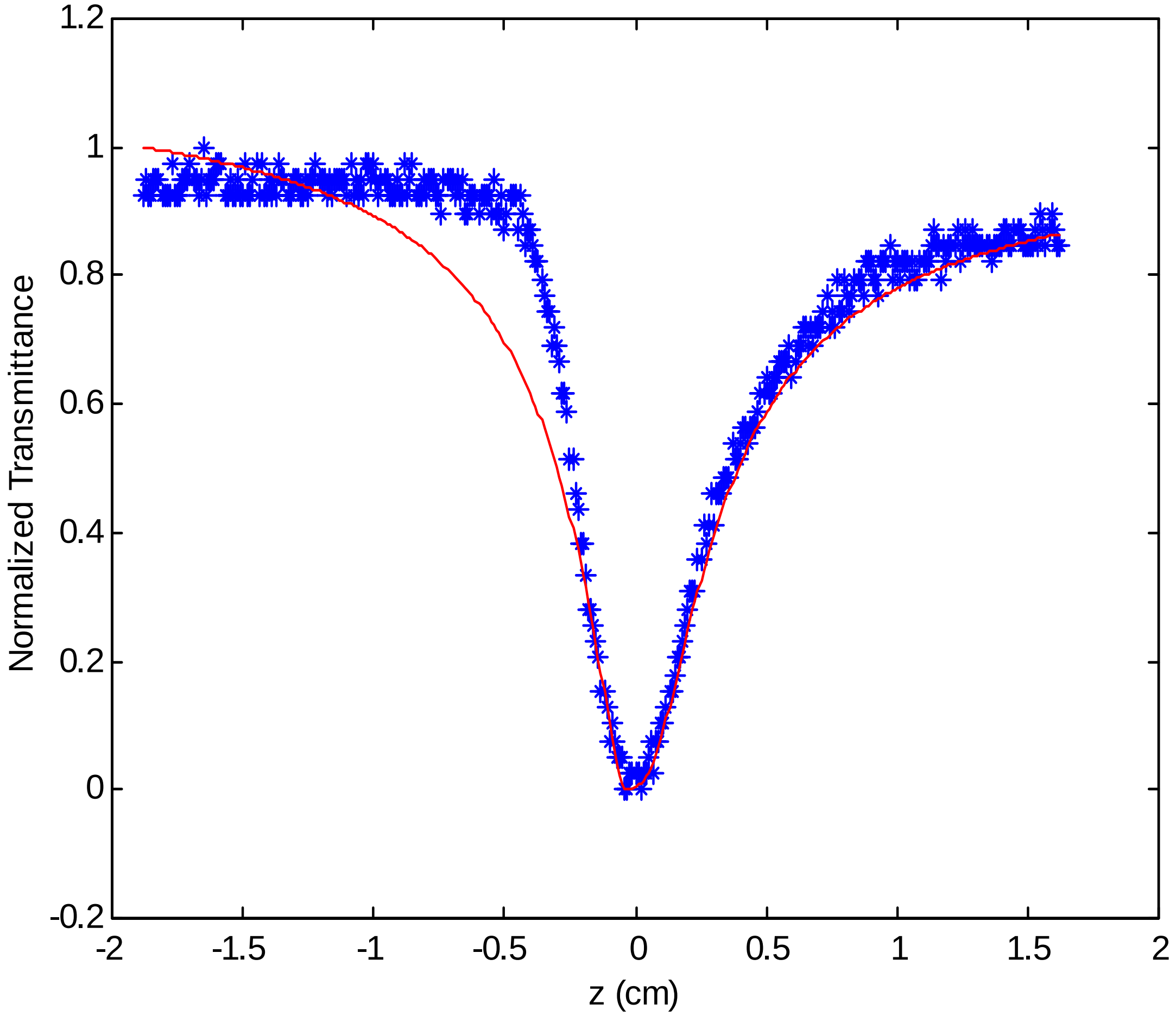

Figure 8c